\documentclass{PoS}

\usepackage{xspace}
\usepackage{txfonts}
\usepackage{multirow}
\usepackage{caption}
\usepackage{setspace}

\title{Heavy Higgs boson resonances and their decay into top quarks at the LHC}

\ShortTitle{Heavy Higgs bosons resonances and their decay into top quarks at the LHC}

\author{Werner Bernreuther$^a$, \speaker{Peter Galler}$^b$,
        Clemens Mellein$^a$, Zong-Guo Si$^c$ and Peter Uwer$^b$\\
	\llap{$^a$}Institut f\"ur Theoretische Teilchenphysik und Kosmologie,\\
        RWTH Aachen University, 52056 Aachen, Germany\\
        \llap{$^b$}Institut f\"ur Physik, Humboldt-Universit\"at zu Berlin,
	12489 Berlin, Germany\\
	\llap{$^c$}School of Physics, Shandong University, Jinan,
	Shandong 250100, China\\
	E-mail: \email{breuther@physik.rwth-aachen.de},
		\email{galler@physik.hu-berlin.de},
		\email{mellein@physik.rwth-aachen.de},
		\email{zgsi@sdu.edu.cn},
		\email{uwer@physik.hu-berlin.de}}

\abstract{
We investigate, within the type-II two-Higgs-doublet extension of the standard
model (SM), the impact of heavy neutral Higgs boson resonances with
unsuppressed Yukawa couplings to top quarks on top-quark pair production at the
LHC at next-to-leading order (NLO) in the strong coupling constant. We take
into account the resonant Higgs boson contributions, the non-resonant SM \ttbar
continuum and the interference of these two contributions. The NLO QCD
corrections to heavy Higgs production and the interference contributions are
calculated in the large top-quark mass ($m_t$) limit, including an effective
K-factor rescaling. Our evaluation of the QCD-Higgs interference is focused
on the Higgs resonance region. Using representative $CP$-conserving as well
as $CP$-violating parameter scenarios phenomenological results are presented
for different observables.
}

\FullConference{Fourth Annual Large Hadron Collider Physics\\
         13-18 June 2016\\
         Lund, Sweden}

\def\ttbar{\ensuremath{t\bar{t}}\xspace}
\def\mtt{\ensuremath{M_{t\bar{t}}}\xspace}
\def\sec#1{Sec.~\ref{#1}}

\def\fig#1{Fig.~\ref{#1}}

\def\eq#1{Eq.~(\ref{#1})}
\def\tab#1{Tab.~\ref{#1}}
\def\vev{{\varv}}
\def\CS{\cos\theta_{\rm{CS}}}

\begin{document}

%
\section{Introduction}
\label{sec:Intro}
After the  125 GeV spin-zero resonance has been experimentally established as a 
Higgs boson with Standard-Model like couplings to fermions and vector
bosons~\cite{Aad:2012tfa,Chatrchyan:2012xdj,Khachatryan:2014jba,Aad:2015gba},
a central research issue at the LHC is the question whether additional
spin-zero bosons exist. We are interested in heavy Higgs bosons with masses
$m_{\phi}>2m_t$ and unsuppressed Yukawa couplings to the top quark that can
appear as resonances in the \ttbar decay channel. Experimental searches
for such resonances have been performed by ATLAS \cite{Aad:2015fna} and
CMS \cite{Chatrchyan:2013lca, Khachatryan:2015sma} but so far no signals above
the SM continuum were detected. However, due to experimental (and theoretical)
uncertainties these measurements are limited in constraining heavy Higgs models.
Furthermore, an important effect in the decay of heavy Higgs bosons to \ttbar
is the interference with the \ttbar continuum that can have a strong influence
on the detectability of these resonances. On the theoretical side the
production of heavy Higgs bosons and decay into top-quark pairs has
been studied until recently only at leading order (LO) in QCD.
NLO corrections as calculated very recently in
\cite{Bernreuther:2015fts,Hespel:2016qaf} have not been taken into account
in the experimental analysis. In this contribution we report on the calculation
\cite{Bernreuther:2015fts} of the NLO QCD corrections to this process,
presenting results for inclusive and differential cross sections at a center
of mass energy of $\sqrt{s}=13$TeV at the LHC. While at LO the full $m_t$
dependence is kept, the NLO corrections are calculated in the heavy top-quark
limit which induces an effective coupling between the Higgs bosons and gluons.
Because relevant contributions from the heavy Higgs bosons to \ttbar production
are only expected in the resonant region, we restricted our computation of the
NLO QCD corrections to this region which simplifies the calculation further.
Our analysis is not restricted to a particular model and can be applied to a
wide spectrum of new physics models involving heavy spin-0 bosons that couple
to \ttbar. However, the choice of a specific model is in order since the decay
widths of the Higgs bosons should be taken into account in a consistent fashion.
This is not only important for preserving unitarity but also because the effects
of heavy Higgs bosons on the \ttbar production sensibly depend on the bosons'
decay widths. Hence, we have chosen a UV-complete theory of an extended Higgs
sector, the type-II two-Higgs-doublet model (2HDM), and the decay widths are
determined by the independent parameters of the model.\\
This article is organized as follows. In \sec{sec:2HDM} we give a brief
overview on aspects of the type-II 2HDM relevant for the following sections.
We study three 2HDM parameter scenarios that are introduced in
\sec{sec:Scenarios}. Leading order results are shown in \sec{sec:LO} in order
to exhibit the main features of heavy Higgs resonances in \ttbar production. In
\sec{sec:NLO} our NLO calculation \cite{Bernreuther:2015fts} is outlined. In
\sec{sec:Results} we show phenomenological results at NLO QCD for different
observables and for the three scenarios introduced in \sec{sec:Scenarios}.
We conclude in \sec{sec:Conclusion}.

%
\section{Type-II two-Higgs-doublet model}
\label{sec:2HDM}
As mentioned in the introduction we choose the type-II two-Higgs-doublet model
as a Higgs sector extension of the SM. It allows us to consistently incorporate
the already discovered SM Higgs boson with $m_h=125$GeV as well as additional
heavy spin-0 bosons which can show up as resonances in the \ttbar decay channel.
Because of strong experimental constraints on flavor changing neutral currents
we choose to study a flavor conserving 2HDM. In particular, we focus on the
\mbox{type-II} 2HDM in which tree-level neutral flavor conservation is realized by
coupling the right-chiral down-type quarks as well as the charged leptons only
to the SU(2)-Higgs doublet $\Phi_1$ and the right-chiral up-type quarks to the
SU(2)-Higgs doublet $\Phi_2$. For a detailed description of the 2HDM we refer
to the literature, e.g. \cite{Branco:2011iw}.\\
The two Higgs doublets are given in the unitary gauge and after electroweak
symmetry breaking by:
\begin{eqnarray} 
\Phi_1 & =& \left(-H^+ \sin\beta,\, \frac{1}{\sqrt{2}}
(\vev_1+\varphi_1-iA\sin\beta)\right)^T, \quad
\Phi_2  =  \left(H^+\cos\beta,\, \, \frac{1}{\sqrt{2}}
(\vev_2+\varphi_2+iA\cos\beta)\right)^T,
\label{eq:2HDM_phunit}
\end{eqnarray}
where $\vev_1$ and $\vev_2$ are the vacuum expectation values of the
two Higgs doublets with $\vev=\sqrt{\vev_1^2+\vev_2^2}=246$ GeV and
$\tan\beta=\vev_2/\vev_1$. The $CP$-even Higgs bosons are denoted by
$\varphi_1$ and $\varphi_2$ and the $CP$-odd Higgs boson is denoted by $A$.
The neutral $CP$-even Higgs bosons can mix with each other and their mass
eigenstates are combinations of $\varphi_1$ and $\varphi_2$ parametrized
by the mixing angle $\alpha$. Furthermore, there is a physical charged Higgs
boson denoted by $H^{\pm}$ which however plays a minor role in our analysis.
The 2HDM also allows for $CP$ violation in the Higgs sector. In this case
the three neutral Higgs bosons can mix with each other such that the
mass eigenstates $\phi_i$ ($i=1,2,3$) are not $CP$ eigenstates. The mixing
is described by an orthogonal mixing matrix $R$ that can be parametrized by
three mixing angles $\alpha_i$ ($i=1,2,3$): 
\begin{equation}
\left(\begin{tabular}{c}
$\phi_1$\\
$\phi_2$\\
$\phi_3$
\end{tabular}\right)=R(\alpha_1,\alpha_2,\alpha_3)\left(
\begin{tabular}{c}
$\varphi_1$\\
$\varphi_2$\\
$A$
\end{tabular}\right).
\end{equation}
For the $CP$-violating 2HDM we choose
\begin{equation}
m_1, m_2, m_3, m_+, \alpha_1, \alpha_2, \alpha_3, \tan\beta, \vev
\end{equation} 
as input parameters. Here $m_i$ ($i=1,2,3$) denote the masses of the neutral
Higgs bosons $\phi_i$ and $m_+$ the mass of $H^+$. For the $CP$ conserving
2HDM the number of independent parameters of the Higgs potential is reduced
and we choose the following set:
\begin{equation}
m_1, m_2, m_3, m_+, \alpha, \tan\beta, \vev,
\end{equation}
where $m_1$ and $m_2$ are the masses of the scalars with $m_1 < m_2$ and
$m_3$ is the mass of the pseudoscalar.

%
\section{Scenarios}
\label{sec:Scenarios}

We investigate three parameter scenarios within the 2HDM, two of them correspond
to $CP$ conservation and one to $CP$ violation by the neutral Higgs-boson
interactions. The parameters are chosen as follows:
\begin{itemize}
\item[a)] The mass of the lightest neutral Higgs boson $\phi_1$ of the model is
put to $m_1=125$ GeV and its couplings are chosen to be equal or close to the
SM Higgs couplings. Equality (i.e., the so-called alignment limit) can be
achieved only in the $CP$-conserving scenarios. In our $CP$-violating scenario
the three neutral Higgs-boson mass eigenstates are $CP$-mixed states. Thus the
couplings of $\phi_1$ differ from the SM Higgs couplings. Yet it is possible to
choose the parameters such that the deviations of the couplings of $\phi_1$ from
the respective SM Higgs couplings are within the experimentally allowed range.
\item[b)] The Yukawa couplings to top quarks are enhanced. In order to estimate
the possible size of the effect in \ttbar production we choose $\tan\beta=0.7$
which leads to top-Yukawa couplings of the neutral Higgs bosons that are still
in accord with experimental constraints. This choice constrains also the mass
of the charged Higgs boson to $m_+>720$ GeV
\cite{Mahmoudi:2009zx,Hermann:2012fc,Eberhardt:2013uba}.
\vspace{-0.1cm}
\item[c)] The masses $m_2,m_3$ of the heavy Higgs bosons are chosen
to be larger than twice the top-quark mass, because we are interested in
the resonant production of $\phi_2,\phi_3$ and their decay to top-quark pairs.
\end{itemize}
The choice of $\tan\beta=0.7$ leads to a suppression of the down-type Yukawa
couplings. Thus in Higgs production by gluon fusion not only the contribution
of the light quarks but also that of the $b$ quark can be safely neglected.\\
In the $CP$-conserving scenarios, the alignment limit is realized in our
parametrization of $R$ by putting $\alpha_1=\beta$. Moreover,
$\alpha_2=\alpha_3=0$. Thus all Yukawa couplings are determined by $\tan\beta$
only. In the alignment limit the coupling of $\phi_1$ to weak vector bosons is
the same as in the SM. Therefore, the couplings of $\phi_2,\phi_3$ to $WW$ and
$ZZ$ are zero. This implies that $\phi_2,\phi_3\rightarrow\ttbar$ is the
dominant decay mode of the heavy Higgs bosons.\\
The masses of the Higgs bosons are free parameters of the 2HDM. The heavy
Higgs boson masses are chosen as follows: In the $CP$-conserving case we
distinguish two scenarios, one with almost degenerate masses (scenario 1)
and the other one with non-degenerate masses (scenario 2). In the $CP$-violating
scenario (scenario 3) the masses of the heavy Higgs bosons are also chosen
to be non-degenerate. The width of the Higgs bosons in the respective scenarios
are fixed by the input parameters and have to be calculated at NLO
(for details see \cite{Bernreuther:2015fts} and references therein).
Table~\ref{tab:scenarios} summarizes the input parameters including the corresponding
decay widths and Yukawa couplings for all three scenarios.
%
%
\begin{table}
\center
{\scriptsize
\begin{tabular}{|c|l||ccc|}
\hline
\multicolumn{2}{|c||}{}& scenario 1 & scenario 2 & scenario 3\\
\hline
\hline
\multirow{9}{*}{\rotatebox[origin=c]{90}{input param.}}
&$\tan\beta$ & 0.7 & 0.7 & 0.7\\
&$\vev$ [GeV] & 246 & 246 & 246\\
&$m_+$ [GeV] & >720& >720 & >720\\
&$m_1$ [GeV] & 125 & 125 & 125\\
&$m_2$ [GeV] & 550 & 550 & 500\\
&$m_3$ [GeV] & 510 & 700 & 800\\
&$\alpha_1$ & $\beta$ & $\beta$ & $\beta$\\
&$\alpha_2$ & 0 & 0 & $\pi/15$\\ 
&$\alpha_3$ & 0 & 0 & $\pi/4$\\ 
\hline
\hline
\multirow{8}{*}{\rotatebox[origin=c]{90}{calculated param.}}
&$\Gamma_2$ [GeV] & 34.56 & 34.49 & 36.55 \\
&$\Gamma_3$ [GeV] & 49.28 & 75.28 & 128.16 \\
&$a_{t1}$ & 1 & 1 & 0.978\\
&$b_{t1}$ & 0 & 0 & 0.297\\
&$a_{t2}$ & 1.429 & 1.429 & 0.863\\
&$b_{t2}$ & 0 & 0 & 0.988\\
&$a_{t3}$ & 0 & 0 & -1.157\\
&$b_{t3}$ & 1.429 & 1.429 & 0.988\\
\hline
\end{tabular}}
\caption{2HDM parameter settings for scenarios 1--3. The SM-like Higgs boson
$\phi_1$ has a decay width $\Gamma_1\approx 4$ MeV. It plays no role in our
analysis.}
\label{tab:scenarios}
\end{table}

%
\section{Leading order results}
\label{sec:LO}

Within the type-II 2HDM we study the resonant production of heavy neutral Higgs
bosons and their decay into top-quark pairs,
\begin{equation}
pp\rightarrow\phi_{2,3}\rightarrow\ttbar X,
\end{equation}
including the SM background process
\begin{equation}
pp\rightarrow\ttbar X,
\end{equation}
as well as the interference of the amplitudes of these two processes. We
investigate this reaction to leading order in the strong coupling constant
$\alpha_s$ in order to illustrate the main features of heavy Higgs resonances
in \ttbar production before we turn to the NLO corrections in the next section.
%
%
\begin{figure}
\begin{centering}
\includegraphics[scale=0.7]{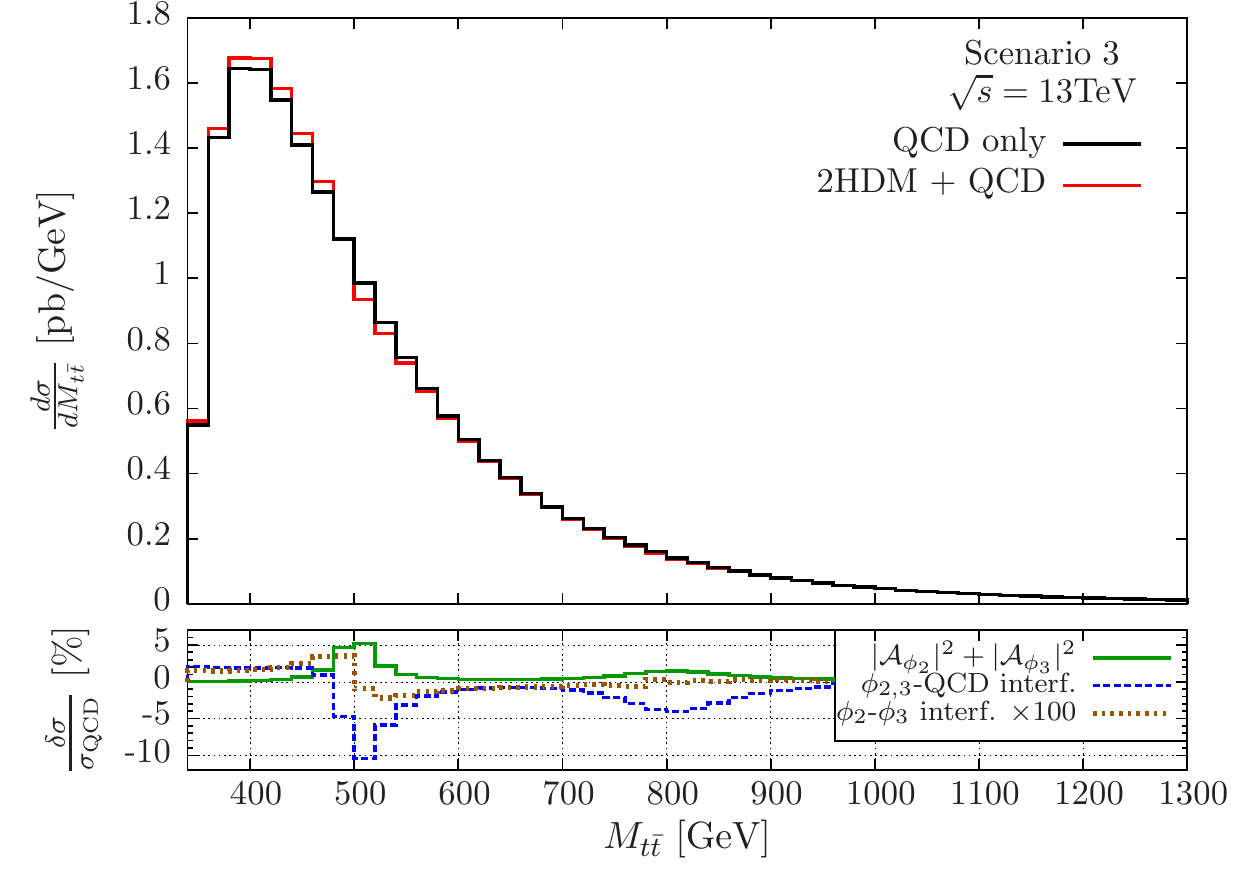}
\caption{Upper plot: \mtt distribution at LO for the QCD background (black)
and the sum of the QCD and Higgs contributions (red) for scenario 3. Lower
plot: Ratio Higgs/QCD for different contributions to the full \mtt distribution.
These are the sum of the squared amplitudes of the two heavy Higgs bosons
(solid, green), the interference with the QCD background (dashed, blue) and the
interference of the two heavy Higgs-boson amplitudes (dotted, brown).}
\label{fig:LOMttbar}
\end{centering}
\end{figure}
%
%
An observable that is sensitive to resonant heavy Higgs bosons in top-quark pair
production is the \ttbar invariant mass \mtt. In \fig{fig:LOMttbar} the \mtt
distribution is shown for \ttbar production by QCD (black) and for the sum of
the QCD and Higgs-boson contributions (red) for scenario 3.
In the lower plot different pieces of the Higgs-boson contributions normalized
to the QCD background are displayed. From this plot one can draw the following
conclusions:
\begin{itemize}
\item The effect of the heavy Higss bosons in \ttbar production is significant 
only in the resonant region. In \fig{fig:LOMttbar} we show the
contribution to the \mtt distribution of the two heavy Higgs bosons within our
2HDM parameter scenario 3 where $m_2 = 500$ GeV and $m_3 = 800$ GeV. The
resonant structures are clearly visible in these regions, especially in the
lower plot, e.g. the green curve that displays the squared amplitudes of the
$\phi_2$ and $\phi_3$ contributions. Away from the resonant region the effects
of the heavy Higgs bosons are very small.
\item The interference of the 2HDM amplitude with the QCD background amplitude
is shown in blue in the lower plot of \fig{fig:LOMttbar}. The interference is
mostly negative and its magnitude is of the same order as the squared Higgs
amplitudes (green curve). This shows that the interference with the background
has a strong effect on the shape of the \mtt distribution and hence can not be
neglected. The sum of the squared Higgs amplitudes and interference term leads
to a peak-dip structure around 500 GeV and a dip around 800 GeV. This fact has
important consequences for the search and detection of a heavy Higgs boson
signal.
\item In the $CP$-violating scenario 3 the Higgs bosons $\phi_j$ are $CP$
mixtures. As a consequence the amplitudes
$gg\rightarrow\phi_j\rightarrow\ttbar$ interfere. This contribution to the
\mtt distribution is shown as brown curve in \fig{fig:LOMttbar}. Note that
in this plot the $\phi_2$-$\phi_3$ interference is multiplied by a factor of
100 for better visibility, i.e. this contribution is strongly suppressed
compared to the other contributions and can safely be neglected. This
suppression is caused by two effects: by the relatively large mass difference
$\Delta m_{23}=m_3-m_2$ and by the choice of parameters in scenario 3, which
leads to cancellations due to opposite-sign Yukawa couplings. (For details see
\cite{Bernreuther:2015fts}.)
\end{itemize}
These observations serve as a guidance for the calculation of the NLO
corrections, which are outlined in the next section. 

%
\section{Next-to-leading order corrections}
\label{sec:NLO}
Because $pp\rightarrow\phi\rightarrow\ttbar$ is already a one-loop process at
LO the NLO corrections involve a two-loop calculation. We have chosen a
simplified approach and computed the NLO corrections to $\phi_j$ ($j=2,3$)
production in the heavy top mass limit which reduces the top-quark loop-induced
$gg\phi$ coupling to a local vertex. This limit can be consistently described
in the context of an effective field theory that yields local effective Higgs
gluon couplings:
\begin{equation}
\mathcal{L}_{\rm{eff}}=\sum_{j=2,3}\left[
 f_{Sj}G_{\mu\nu}^aG^{\mu\nu}_a
+f_{Pj}\epsilon_{\mu\nu\alpha\beta}G^{\mu\nu}_aG^{\alpha\beta}_a
\right]\phi_j
\label{eq:Leff}
\end{equation}
where $f_{Sj}$ and $f_{Pj}$ are the Wilson coefficients describing the coupling
of the scalar and pseudoscalar component of the Higgs boson $\phi_j$ to gluons.
The effective Lagrangian in \eq{eq:Leff} also contains couplings of $\phi_j$ to
three and four gluons, of which only the former is relevant for heavy Higgs
production and decay to \ttbar at NLO. In the heavy top-mass limit the two-loop
calculation becomes effectively a one-loop calculation. This limit can also be
understood as the first order in an expansion in $m_{\phi}/m_t$. In the case of
Higgs production it has been shown \cite{Kramer:1996iq} that even for
$m_{\phi} > m_t$ this approach gives results with rather small uncertainties of
about 10\% depending on the size of $m_{\phi}$ if an appropriate $K$ factor is
applied to the NLO corrections in the effective theory. We apply this procedure
in the following way in order to get an approximate result for the cross section
of our process at NLO QCD:
\begin{equation}
\sigma_{\rm{NLO}}^{\rm{approx}}\equiv\sum_{j=2,3}\left(
\sigma_{{\rm full},j}^{(0)}+\sigma_{{\rm full},j,\rm{QCD}}^{(0)}
+K_j\sigma_{{\rm eff},j}^{(1)}+\sigma_{{\rm eff},j,\rm{QCD}}^{(1)}
\right)\quad\mbox{with}\quad
K_j=\frac{\sigma_{{\rm full},j}^{(0)}}{\sigma_{{\rm eff},j}^{(0)}},
\label{Kfactorformula}
\end{equation}
Here $\sigma_{{\rm full},j}^{(0)}$ denote the LO cross sections for
\begin{equation}
pp\rightarrow\phi_j\rightarrow\ttbar
\label{phiprocess}
\end{equation}
and $\sigma_{{\rm full},\rm{QCD}}^{(0)}$ results from the interference of the
amplitude of (\ref{phiprocess}) and of the QCD background at LO. The subscript \textit{full}
indicates that the full top-quark mass dependence is kept. The terms
$\sigma_{{\rm eff},j}^{(1)}$ ($\sigma_{{\rm eff},j}^{(0)}$) and 
$\sigma_{{\rm eff},j,\rm{QCD}}^{(1)}$ represent the NLO (LO) cross sections
for the process (\ref{phiprocess}) and the interference with the QCD background
at NLO, respectively, in the effective theory. While \eq{Kfactorformula} is
given for inclusive cross sections an analogous formula can also be applied to
individual bins of differential distributions.\\
In addition to the heavy top limit we apply a second approximation.
As the LO results in \sec{sec:LO} show, the dominant contribution from
heavy Higgs bosons to \ttbar production comes from the resonant region. Hence
we restrict the NLO calculation of the differential distributions to this region.
The calculation involves factorizable and non-factorizable QCD corrections to LO
Higgs production and decay. The former are manifestly resonant while the latter
consist of resonant and non-resonant contributions. The resonant parts of the
non-factorizable contributions can be extracted by applying the soft gluon approximation
\cite{Fadin:1993dz,Melnikov:1995fx,Beenakker:1997ir,Dittmaier:2014qza}.
Within this approximation non-factorizable contributions from the
real and virtual corrections cancel exactly. Thus one is left with only
factorizable contributions. (For details see \cite{Bernreuther:2015fts}.)\\
Apart from the two approximations we apply standard techniques to perform
the NLO calculation. In particular, we use the dipole subtraction method
\cite{Catani:1996vz,Catani:2002hc} to handle the infrared divergences in
the real and virtual corrections.
Analytical results of our NLO calculation of the squared matrix elements are
given in \cite{Bernreuther:2015fts}.

%
\section{Results}
\label{sec:Results}
\begin{figure}
\includegraphics[scale=0.6]{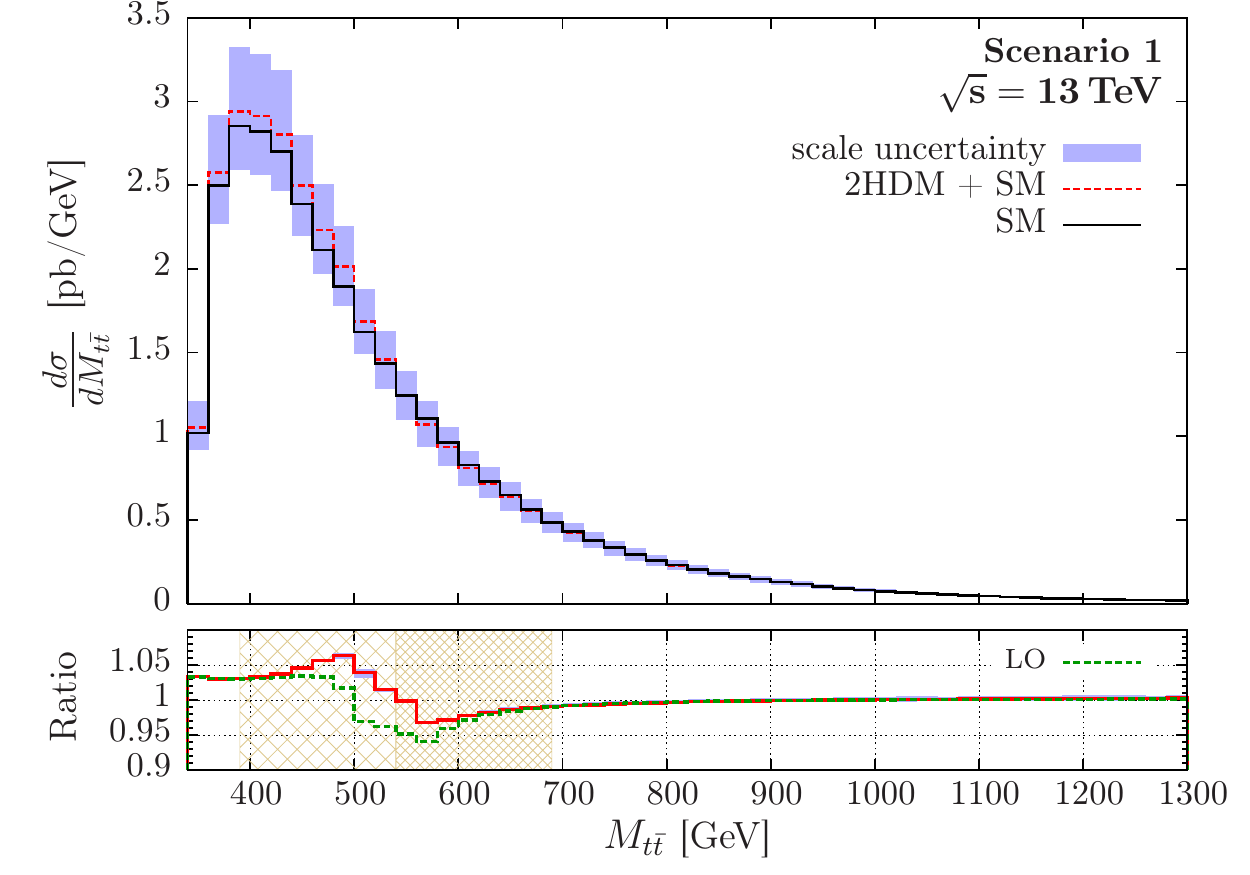}
\includegraphics[scale=0.6]{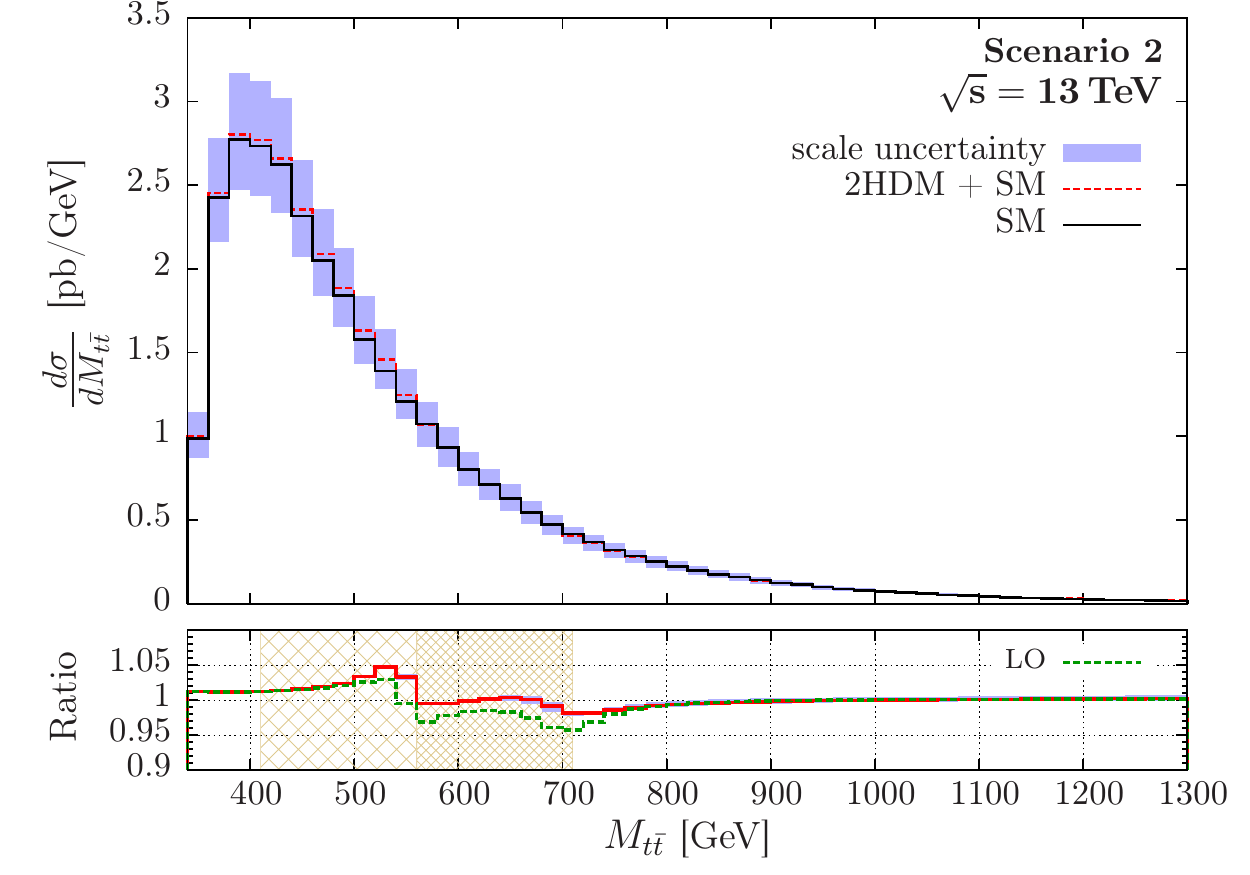}
\begin{centering}
\includegraphics[scale=0.6]{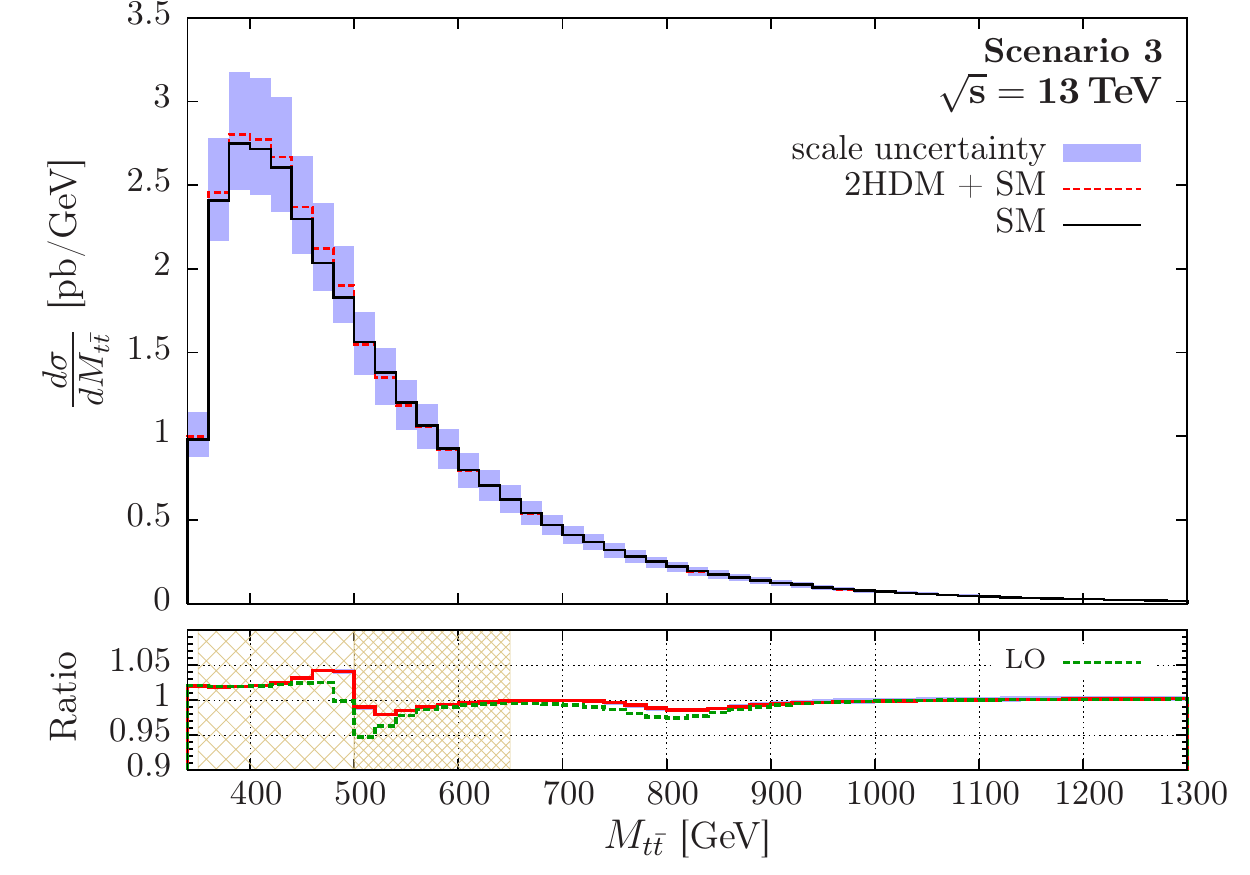}
\end{centering}
\caption{\mtt distribution for scenario 1--3. Upper left: scenario 1,
upper right: scenario 2, lower plot: scenario 3.}
\label{fig:Mtt}
\end{figure}

\begin{figure}
\centering
\begin{minipage}[t]{.5\textwidth}
  \includegraphics[scale=0.6]{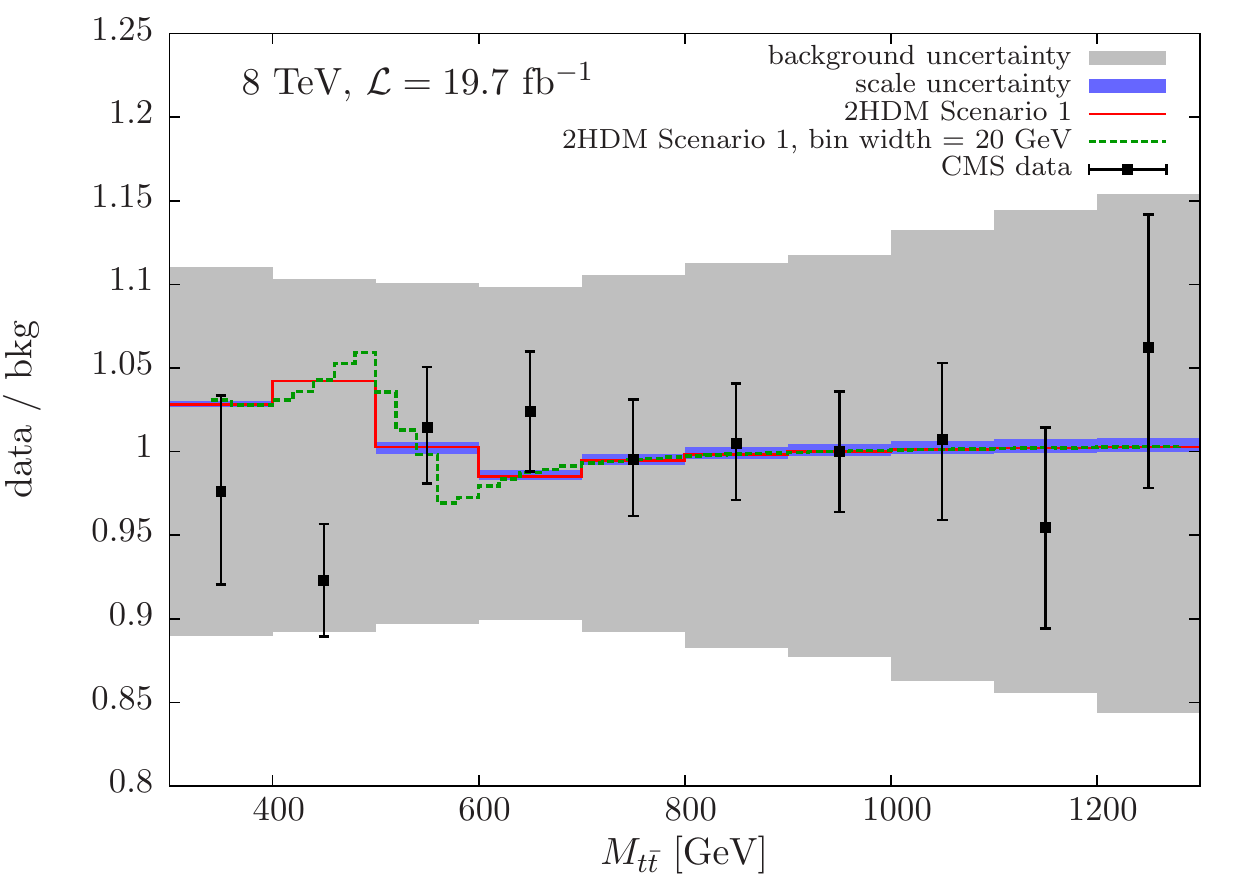}
  \captionsetup{width=0.9\linewidth}
  \captionof{figure}{CMS result for the ratio data/bkg. binned in \mtt at $\sqrt{s}=8$ TeV
taken from \cite{Chatrchyan:2013lca}. The theoretical prediction for scenario 1
is plotted in red (100 GeV binning) and green (20 GeV binning).}
  \label{fig:CMS}
\end{minipage}%
\begin{minipage}[t]{.5\textwidth}
  \includegraphics[scale=0.6]{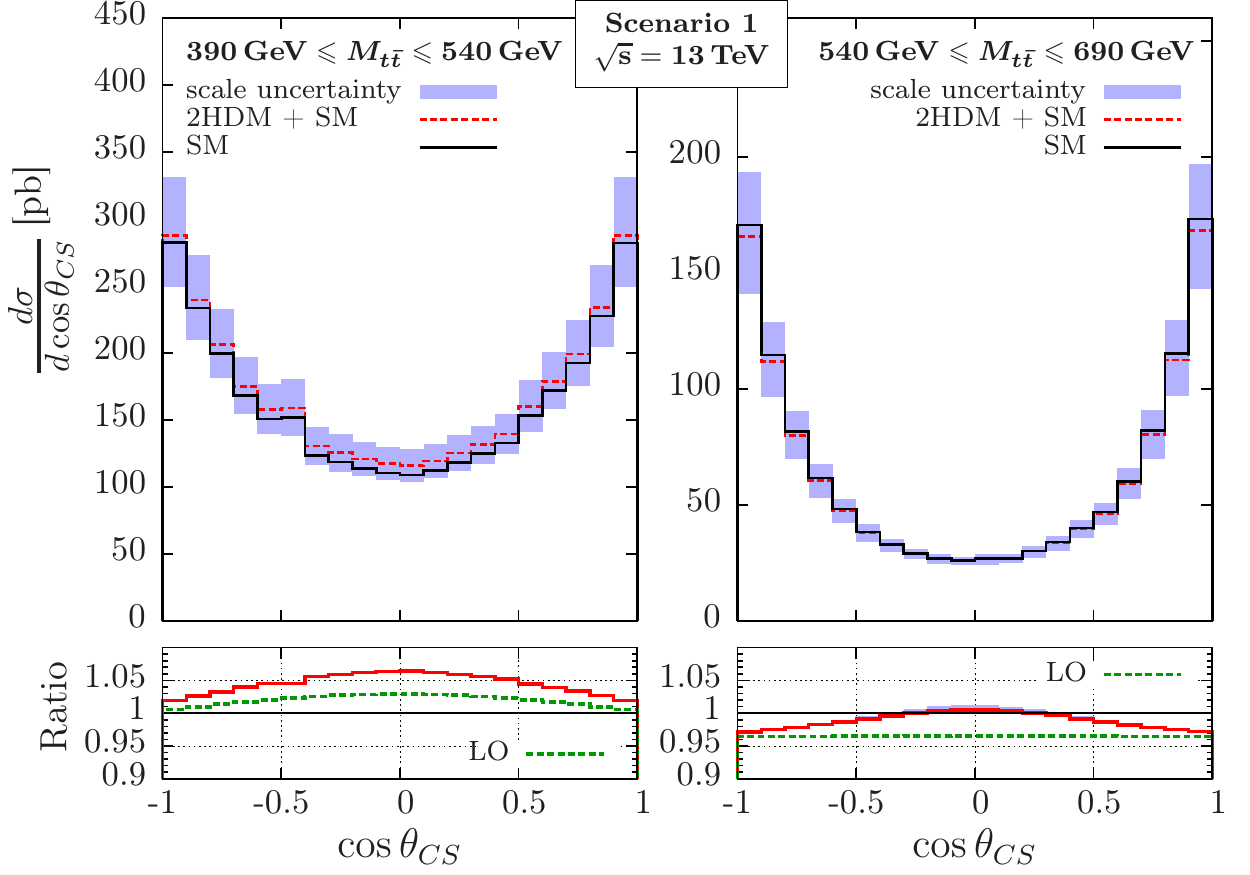}
  \captionsetup{width=0.9\linewidth}
  \captionof{figure}{Distribution of the Collins-Soper angle for scenario 1 with two different
  \mtt cuts.}
  \label{fig:CS}
\end{minipage}
\end{figure}

In this section we present the results for heavy Higgs production and decay
to \ttbar including the approximate NLO QCD corrections to the 2HDM and
2HDM-QCD interference contributions, as well as the full NLO QCD and weak
interaction corrections to the background.\\
The inclusive cross sections for \ttbar production are listed in
\tab{tab:inclxs} for the three parameter scenarios. The label 2HDM denotes the
contribution to the inclusive \ttbar cross section from the two heavy Higgs
bosons and the interference with the QCD background at NLO using the
aforementioned approximations, while $\sigma_{\rm{QCDW}}$ is the contribution
of the QCD background at NLO including the weak interaction corrections.
The ratio $\sigma_{\rm{2HDM}}/\sigma_{\rm{QCDW}}$ is rather small; the 2HDM
contribution is about 2\% of the inclusive QCD cross section in the case of
scenario 1 and about 1\% for scenarios 2 and 3. There are mainly two effects
causing the inclusive cross section to be hardly sensitive to the heavy Higgs
resonances. The QCD-induced \ttbar production is large and the interference
effects partly cancel the resonant enhancement of the Higgs contributions. It
is therefore important to study other, more sensitive observables, in
particular observables evaluated in \ttbar invariant mass bins.\\
%
%
\begin{table}[h!]
\centering
\begin{tabular}{l|lll}
& Scenario 1 & Scenario 2 & Scenario 3\\
\hline
$\mu_0$ [GeV]& 265 & 312.5 & 325\\
$\sigma_{\rm{QCDW}}$ [pb] & $643.22^{+81.23}_{-77.71}$&
$624.25^{+80.98}_{-76.19}$& $619.56^{+81.05}_{-75.72}$\\
$\sigma_{\rm{2HDM}}$ [pb] & $13.59^{+1.85}_{-1.64}$&
$7.4^{+0.77}_{-0.78}$& $7.21^{+0.81}_{-0.77}$\\
\hline
$\sigma_{\rm{2HDM}}/\sigma_{\rm{QCDW}}$ [\%]& 2.1& 1.2& 1.2\end{tabular}
\caption{Inclusive \ttbar cross sections without and with the heavy Higgs
resonances.}
\label{tab:inclxs}
\end{table}
%
%
The \ttbar invariant mass distribution $d\sigma/d\mtt$ is shown
in \fig{fig:Mtt} for scenarios 1--3. 
The upper plots display the SM \mtt distribution and the sum of SM and 2HDM
contribution (``squared'' and interference) at NLO including the 
uncertainty due to renormalization and factorization scale variations.
The lower plots in \fig{fig:Mtt} show the ratio (QCDW + 2HDM)/QCDW at NLO (red)
and at LO (green). The strongest effect of about 7\% appears in those bins
of the \mtt distribution in scenario 1 where the contributions of the two
resonances overlap.
%
%
%
%
In \fig{fig:CMS} we show a comparison of our result for the \mtt distribution
within scenario 1 with a CMS analysis~\cite{Chatrchyan:2013lca} at 8 TeV
center-of-mass energy using 19.7 fb$^{-1}$ of data.
As can be seen from \fig{fig:CMS} this analysis is not yet sensitive enough
to constrain scenario 1. A similar analysis from ATLAS exists\footnote{
The ATLAS analysis \cite{Aad:2015fna} of the \mtt distribution at 8 TeV is
also not sensitive enough to constrain this model because the interference
effects were not taken into account. However, during the writeup of this
proceedings contribution an updated ATLAS analysis \cite{ATLAS:2016pyq}
was published. This analysis is considerably more sensitive for several reasons.
It succeeds in reducing the uncertainty on the background, it uses a
smaller \mtt bin width and takes interference effects into account.}
\cite{Aad:2015fna} with a similar sensitivity.
In \fig{fig:CMS} one can see that the background uncertainty of the
analysis~\cite{Chatrchyan:2013lca} is too large to detect the signal.
Furthermore the large bin size of 100 GeV causes the theoretical
signal-to-background ratio to fall from a peak value of about 7\% to
below 5\%.\\
%
%
%
%
By imposing cuts on \mtt (indicated by hatched regions in \fig{fig:Mtt})
below and above the resonance we avoid the peak-dip
cancellation and obtain an estimate on the largest possible Higgs effects in
different observables. Besides the \mtt distribution we studied the
distribution of the top-quark rapidity, of the top-quark transverse
momentum, and of the Collins-Soper angle $\theta_{\rm{CS}}$. Among these
observables the distribution $d\sigma/d\CS$ is the most sensitive one
to a heavy Higgs boson. This distribution is shown in \fig{fig:CS} for
scenario 1 for two different \mtt cuts. The lower plots show the
signal-to-background ratio which is $>5\%$ between $\CS=-0.4$ and $\CS=0.5$
in the lower \mtt bin. Furthermore, \fig{fig:CS} illustrates the importance
of the NLO corrections which lead to an enhancement of the signal-to-background
ratio.

%
\section{Conclusion}
\label{sec:Conclusion}
We calculated the NLO QCD corrections to the resonant production of heavy Higgs
bosons at the LHC and their decay into top-quark pairs in the large top mass
limit. We have taken into account the SM $\ttbar$ continuum and its
interference with the signal at NLO. This interference has an important
effect on the shape of the $\ttbar$ invariant mass distribution, $\mtt$, in
the vicinity of a heavy Higgs resonance. Depending on the mass and the
couplings of the resonance it leads to a peak-dip structure or to a dip rather
than a bump in the $\mtt$ spectrum. We found that the QCD corrections to these
$\mtt$ shape distortions, but also to the distribution of the Collins-Soper
angle, are significant in the resonance region. We investigated these heavy
Higgs-boson effects within a type-II two-Higgs-doublet extension of the SM.\\
The shape-distortions of the $\mtt$ spectrum predicted at NLO QCD have
important consequences for the search for a heavy Higgs-boson signal in the
measured $\ttbar$ invariant mass distribution. If this distribution is
measured with a large bin size the peak and dip structure caused by a Higgs
resonance averages out. As a consequence the inclusive \ttbar cross section
is affected by the heavy Higgs-boson contributions in our scenarios only at
the level of 1-2\%. A dedicated resonance search could be done by scanning
the measured $\ttbar$ invariant mass spectrum with a sliding $\mtt$ window
with the smallest experimentally attainable width.

\begin{spacing}{0}
\bibliographystyle{JHEP}
\bibliography{references}

\providecommand{\href}[2]{#2}\begingroup\raggedright\begin{thebibliography}{10}

\bibitem{Aad:2012tfa}
{\scshape ATLAS} collaboration, G.~Aad et~al., \emph{{Observation of a new
  particle in the search for the Standard Model Higgs boson with the ATLAS
  detector at the LHC}},
  \href{http://dx.doi.org/10.1016/j.physletb.2012.08.020}{\emph{Phys. Lett.}
  {\bf B716} (2012) 1--29}, [\href{http://arxiv.org/abs/1207.7214}{{\tt
  1207.7214}}].

\bibitem{Chatrchyan:2012xdj}
{\scshape CMS} collaboration, S.~Chatrchyan et~al., \emph{{Observation of a new
  boson at a mass of 125 GeV with the CMS experiment at the LHC}},
  \href{http://dx.doi.org/10.1016/j.physletb.2012.08.021}{\emph{Phys. Lett.}
  {\bf B716} (2012) 30--61}, [\href{http://arxiv.org/abs/1207.7235}{{\tt
  1207.7235}}].

\bibitem{Khachatryan:2014jba}
{\scshape CMS} collaboration, V.~Khachatryan et~al., \emph{{Precise
  determination of the mass of the Higgs boson and tests of compatibility of
  its couplings with the standard model predictions using proton collisions at
  7 and 8 TeV}},
  \href{http://dx.doi.org/10.1140/epjc/s10052-015-3351-7}{\emph{Eur. Phys. J.}
  {\bf C75} (2015) 212}, [\href{http://arxiv.org/abs/1412.8662}{{\tt
  1412.8662}}].

\bibitem{Aad:2015gba}
{\scshape ATLAS} collaboration, G.~Aad et~al., \emph{{Measurements of the Higgs
  boson production and decay rates and coupling strengths using pp collision
  data at $\sqrt{s}=7$ and 8 TeV in the ATLAS experiment}},
  \href{http://dx.doi.org/10.1140/epjc/s10052-015-3769-y}{\emph{Eur. Phys. J.}
  {\bf C76} (2016) 6}, [\href{http://arxiv.org/abs/1507.04548}{{\tt
  1507.04548}}].

\bibitem{Aad:2015fna}
{\scshape ATLAS} collaboration, G.~Aad et~al., \emph{{A search for $
  t\overline{t} $ resonances using lepton-plus-jets events in proton-proton
  collisions at $ \sqrt{s}=8 $ TeV with the ATLAS detector}},
  \href{http://dx.doi.org/10.1007/JHEP08(2015)148}{\emph{JHEP} {\bf 08} (2015)
  148}, [\href{http://arxiv.org/abs/1505.07018}{{\tt 1505.07018}}].

\bibitem{Chatrchyan:2013lca}
{\scshape CMS} collaboration, S.~Chatrchyan et~al., \emph{{Searches for new
  physics using the $t\bar{t}$ invariant mass distribution in pp collisions at
  $\sqrt{s}=8$ TeV}}, \href{http://dx.doi.org/10.1103/PhysRevLett.111.211804,
  10.1103/PhysRevLett.112.119903}{\emph{Phys. Rev. Lett.} {\bf 111} (2013)
  211804}, [\href{http://arxiv.org/abs/1309.2030}{{\tt 1309.2030}}].

\bibitem{Khachatryan:2015sma}
{\scshape CMS} collaboration, V.~Khachatryan et~al., \emph{{Search for resonant
  $t \bar t$ production in proton-proton collisions at $\sqrt{s}=8$ TeV}},
  \href{http://dx.doi.org/10.1103/PhysRevD.93.012001}{\emph{Phys. Rev.} {\bf
  D93} (2016) 012001}, [\href{http://arxiv.org/abs/1506.03062}{{\tt
  1506.03062}}].

\bibitem{Bernreuther:2015fts}
W.~Bernreuther, P.~Galler, C.~Mellein, Z.~G. Si and P.~Uwer, \emph{{Production
  of heavy Higgs bosons and decay into top quarks at the LHC}},
  \href{http://dx.doi.org/10.1103/PhysRevD.93.034032}{\emph{Phys. Rev.} {\bf
  D93} (2016) 034032}, [\href{http://arxiv.org/abs/1511.05584}{{\tt
  1511.05584}}].

\bibitem{Hespel:2016qaf}
B.~Hespel, F.~Maltoni and E.~Vryonidou, \emph{{Signal background interference
  effects in heavy scalar production and decay to a top-anti-top pair}},
  \href{http://arxiv.org/abs/1606.04149}{{\tt 1606.04149}}.

\bibitem{Branco:2011iw}
G.~C. Branco, P.~M. Ferreira, L.~Lavoura, M.~N. Rebelo, M.~Sher and J.~P.
  Silva, \emph{{Theory and phenomenology of two-Higgs-doublet models}},
  \href{http://dx.doi.org/10.1016/j.physrep.2012.02.002}{\emph{Phys. Rept.}
  {\bf 516} (2012) 1--102}, [\href{http://arxiv.org/abs/1106.0034}{{\tt
  1106.0034}}].

\bibitem{Mahmoudi:2009zx}
F.~Mahmoudi and O.~Stal, \emph{{Flavor constraints on the two-Higgs-doublet
  model with general Yukawa couplings}},
  \href{http://dx.doi.org/10.1103/PhysRevD.81.035016}{\emph{Phys. Rev.} {\bf
  D81} (2010) 035016}, [\href{http://arxiv.org/abs/0907.1791}{{\tt
  0907.1791}}].

\bibitem{Hermann:2012fc}
T.~Hermann, M.~Misiak and M.~Steinhauser, \emph{{$\bar{B}\to X_s \gamma$ in the
  Two Higgs Doublet Model up to Next-to-Next-to-Leading Order in QCD}},
  \href{http://dx.doi.org/10.1007/JHEP11(2012)036}{\emph{JHEP} {\bf 11} (2012)
  036}, [\href{http://arxiv.org/abs/1208.2788}{{\tt 1208.2788}}].

\bibitem{Eberhardt:2013uba}
O.~Eberhardt, U.~Nierste and M.~Wiebusch, \emph{{Status of the
  two-Higgs-doublet model of type II}},
  \href{http://dx.doi.org/10.1007/JHEP07(2013)118}{\emph{JHEP} {\bf 07} (2013)
  118}, [\href{http://arxiv.org/abs/1305.1649}{{\tt 1305.1649}}].

\bibitem{Kramer:1996iq}
M.~Kramer, E.~Laenen and M.~Spira, \emph{{Soft gluon radiation in Higgs boson
  production at the LHC}},
  \href{http://dx.doi.org/10.1016/S0550-3213(97)00679-2}{\emph{Nucl. Phys.}
  {\bf B511} (1998) 523--549}, [\href{http://arxiv.org/abs/hep-ph/9611272}{{\tt
  hep-ph/9611272}}].

\bibitem{Fadin:1993dz}
V.~S. Fadin, V.~A. Khoze and A.~D. Martin, \emph{{Interference radiative
  phenomena in the production of heavy unstable particles}},
  \href{http://dx.doi.org/10.1103/PhysRevD.49.2247}{\emph{Phys. Rev.} {\bf D49}
  (1994) 2247--2256}.

\bibitem{Melnikov:1995fx}
K.~Melnikov and O.~I. Yakovlev, \emph{{Final state interaction in the
  production of heavy unstable particles}},
  \href{http://dx.doi.org/10.1016/0550-3213(96)00151-4}{\emph{Nucl. Phys.} {\bf
  B471} (1996) 90--120}, [\href{http://arxiv.org/abs/hep-ph/9501358}{{\tt
  hep-ph/9501358}}].

\bibitem{Beenakker:1997ir}
W.~Beenakker, A.~P. Chapovsky and F.~A. Berends, \emph{{Nonfactorizable
  corrections to W pair production: Methods and analytic results}},
  \href{http://dx.doi.org/10.1016/S0550-3213(97)00628-7}{\emph{Nucl. Phys.}
  {\bf B508} (1997) 17--63}, [\href{http://arxiv.org/abs/hep-ph/9707326}{{\tt
  hep-ph/9707326}}].

\bibitem{Dittmaier:2014qza}
S.~Dittmaier, A.~Huss and C.~Schwinn, \emph{{Mixed QCD-electroweak
  $\mathcal{O}(\alpha_s\alpha)$ corrections to Drell-Yan processes in the
  resonance region: pole approximation and non-factorizable corrections}},
  \href{http://dx.doi.org/10.1016/j.nuclphysb.2014.05.027}{\emph{Nucl. Phys.}
  {\bf B885} (2014) 318--372}, [\href{http://arxiv.org/abs/1403.3216}{{\tt
  1403.3216}}].

\bibitem{Catani:1996vz}
S.~Catani and M.~H. Seymour, \emph{{A General algorithm for calculating jet
  cross-sections in NLO QCD}},
  \href{http://dx.doi.org/10.1016/S0550-3213(96)00589-5}{\emph{Nucl. Phys.}
  {\bf B485} (1997) 291--419}, [\href{http://arxiv.org/abs/hep-ph/9605323}{{\tt
  hep-ph/9605323}}].

\bibitem{Catani:2002hc}
S.~Catani, S.~Dittmaier, M.~H. Seymour and Z.~Trocsanyi, \emph{{The Dipole
  formalism for next-to-leading order QCD calculations with massive partons}},
  \href{http://dx.doi.org/10.1016/S0550-3213(02)00098-6}{\emph{Nucl. Phys.}
  {\bf B627} (2002) 189--265}, [\href{http://arxiv.org/abs/hep-ph/0201036}{{\tt
  hep-ph/0201036}}].

\bibitem{ATLAS:2016pyq}
{\scshape ATLAS} collaboration, \emph{{Search for heavy Higgs bosons A/H
  decaying to a top-quark pair in pp collisions at $\sqrt{s}=8$ TeV with the
  ATLAS detector}}, ATLAS-CONF-2016-073.

\end{thebibliography}\endgroup
\end{spacing}
\end{document}